\documentclass[review,dvipdfmx]{elsarticle}
\usepackage{amsmath}
\usepackage{amssymb}
\usepackage{bm}
\usepackage{algorithmic}
\usepackage{algorithm2e}
\usepackage{color}
\usepackage{verbatim}
\usepackage{listings}

\journal{Computer Physics Communications}

\begin{document}

\title{A generalized GPU-based connected component labeling algorithm} 

\author[a]{Yukihiro Komura\corref{author}}
\ead{yukihiro.komura.ss@alum.riken.jp}
\cortext[author] {Corresponding author.} 
\address[a]{RIKEN, Advanced Institute for Computational Science, 
7-1-26 Minatojima-minami-machi, Chuo-ku, Kobe, Hyogo 650-0047, Japan
}

\begin{abstract}
We propose a generalized GPU-based connected component labeling (CCL) algorithm 
that can be applied to both various lattices and to non-lattice environments 
in a uniform fashion. 
We extend our recent GPU-based CCL algorithm without the use of conventional iteration to the generalized method. 
As an application of this algorithm, we deal with the bond percolation problem. 
We investigate bond percolation on the honeycomb and 
triangle lattices to confirm the correctness of this algorithm. 
Moreover, we deal with bond percolation on 
the Bethe lattice as a substitute for a network structure, and demonstrate 
the performance of this algorithm on those lattices. 
\end{abstract}


\maketitle

\section{Introduction}
The connected component labeling (CCL) algorithm has been used 
to study image processing \cite{Suzuki00,Suzuki09,Hawick_labeling,Kalentev}, 
applications in physics \cite{komura12, komura14, komura15, komura15_program}, percolation theory, 
and a problem involving porous rocks \cite{Futaisia03}. 
The majority of CCL algorithms were designed for lattice environments, with 
only a few studies of CCL algorithms for non-lattice environments. 
In non-lattice environments, the positions of the sites 
are arbitrary rather than being restricted to discrete points of a regular lattice. 
Non-lattice environments exist not only in the percolation theory of 
disordered discs and spheres, but also in networks.

The performance of a single CPU core remains almost unchanged from a decade ago. 
However, the number of cores has increased each year, and 
recent advances in application performance
have been realized by exploiting such concepts as multiple cores and many threads. 
Graphics accelerators
are a common example of many-thread devices, having
evolved into highly parallel threads 
with very high memory bandwidth, and
research has clearly shown that they can dramatically improve computing performance.  
The most widely used programming model for accelerators is CUDA \cite{cuda}, 
a parallel-computing platform and programming model developed by NVIDIA 
that is essentially C/C++ with several extensions that allow functions to be executed directly on the NVIDIA GPU. 

Connected component labeling algorithms with a single GPU have been proposed by many researchers. 
The majority of those CCL algorithms have been designed for lattice environments, and  
use a two-stage approach 
that divides the lattice into sub-blocks that are independently treated and then merged. 
However, the sub-block decomposition is difficult in non-lattice environments.
Thus we need GPU-based CCL algorithms without a sub-block decomposition 
for computation in non-lattice environments. 

Here we describe several experiments on CCL algorithms with a single GPU   
and without the sub-block decomposition.
Hawick {\it et al.} \cite{Hawick_labeling} proposed a 
CCL algorithm called ''Label Equivalence'', and 
Kalentev {\it et al.} \cite{Kalentev} improved the label equivalence algorithm. 
Both of these algorithms are realized by an iterative method  
of comparison with nearest-neighbor sites. 
More recently, the present author \cite{komura15} has also proposed a single GPU 
CCL algorithm that does not use conventional iteration. 
The computation times using this approach have proved to be 
about half those of the previous method \cite{Kalentev} for 
the application of the Swendsen-Wang multi-cluster spin-flip algorithm \cite{sw87}. 

The above described CCL algorithms focus on the case of square and simple cubic lattices, 
and the algorithm in \cite{komura15} is specialized to the case of square and simple cubic lattices. 
In this paper, we propose a generalized GPU-based CCL algorithm 
that can be applied to both various lattices and to non-lattice environments in a uniform fashion. 
This generalized method extends our recent GPU-based CCL algorithm \cite{komura15}, which does not use 
conventional iteration. 
To confirm the correctness and performance of this algorithm, 
we deal with bond percolation problems.
This paper is organized as follows: 
In Section~2, we briefly review the GPU-based CCL algorithm that serves as a starting point 
\cite{komura15}.
In Section~3, we describe the new generalized algorithm. 
In Section~4, we first show the result of 
the bond percolations for the honeycomb and 
triangle lattices to confirm the correctness.  
Moreover, we adapt the generalized algorithm to 
bond percolation for the Bethe lattice as a substitute for a network structure, 
and the resulting performance is described. 
Finally, a summary and discussion are presented in Section~5.

\section{Connected Component Labeling}
Connected component labelling algorithms assign proper cluster labels to each site
on the basis of local connection information. 
Many CCL algorithms have been proposed by many researchers. 
The Hoshen--Kopelman algorithm \cite{Hoshen_Kopelman} and 
the CCL algorithm proposed by Suzuki {\it et al.} \cite{Suzuki00} 
are the best-known CCL algorithms that use a single CPU. 
The majority of CCL algorithms using a single CPU 
are specialized to the case of square and simple cubic lattices. 
Moreover, those algorithms are realized by sequential computation. 
However, sequential algorithms cannot be applied to GPU computation, 
and thus many CCL algorithms that use a single CPU cannot be directly applied to GPU computation, 
so a suitable new algorithm is required. 

We briefly review the CCL algorithm without conventional iteration~\cite{komura15}. 
The cornerstone of this algorithm
is the method of label reduction proposed by Wende et al.~\cite{Wende13}; 
the procedure is illustrated in Fig.~1 of \cite{komura15}. 
The labeling consists of four steps:
(i) initialization, (ii) analysis, 
(iii) label reduction, and (iv) analysis. 
This method does not require an iterative method of comparison with nearest-neighbor sites. The number of such comparisons in this method 
is 1 for a square lattice and 2 for a simple cubic lattice if periodic boundary conditions are not used.
In the initialization function, 
a label is assigned to each site based on its connections: 
each site has $\textit{label}\mskip 2mu [i]=\textit{min}$, where \textit{min} is the
lowest-numbered of the connected sites.
The analysis function tracks the label
from a given site to a new site determined by the value of the label at the given site. 
All sites have $\textit{label}\mskip 2mu [i]=\textit{label}\mskip 2mu [\textit{label}\mskip 2mu [i]]$ calculated until \textit{label} remains unchanged. 
In the label reduction step, we use Algorithm~1 of \cite{komura15} on all sites. 
The reduction method is used in only 
one direction for square lattices and in only two directions for simple cubic lattices. 
To resolve conflicts in the label update process, 
the atomic function is used in Algorithm~1 of \cite{komura15}, 
and each chain of labels, i.e., $\textit{label}\mskip 2mu [\textit{label}\mskip 2mu ]$, is constructed automatically. 
Finally, the analysis function is executed again. 
The cluster labeling algorithm in \cite{komura15} uses a \texttt{while} loop  
within the label reduction step. 
However, the number of sites for which the \texttt{while} loop is executed in this step 
is kept to a minimum.

\begin{figure}
\begin{center}
\includegraphics[width=12.0cm]{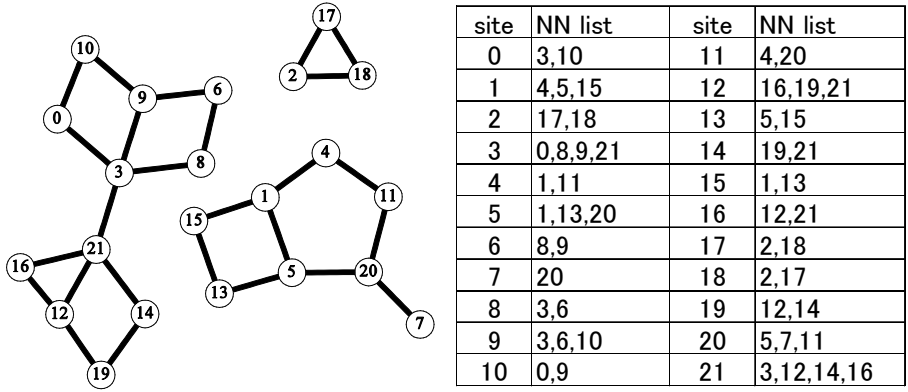}
\caption{\label{fig:method1} 
Example of the nearest-neighbor (NN) list for a complex network. 
The number at each site is the index, and each has a NN list.  
}
\end{center}
\end{figure}

\begin{figure}
\begin{center}
\includegraphics[width=12.0cm]{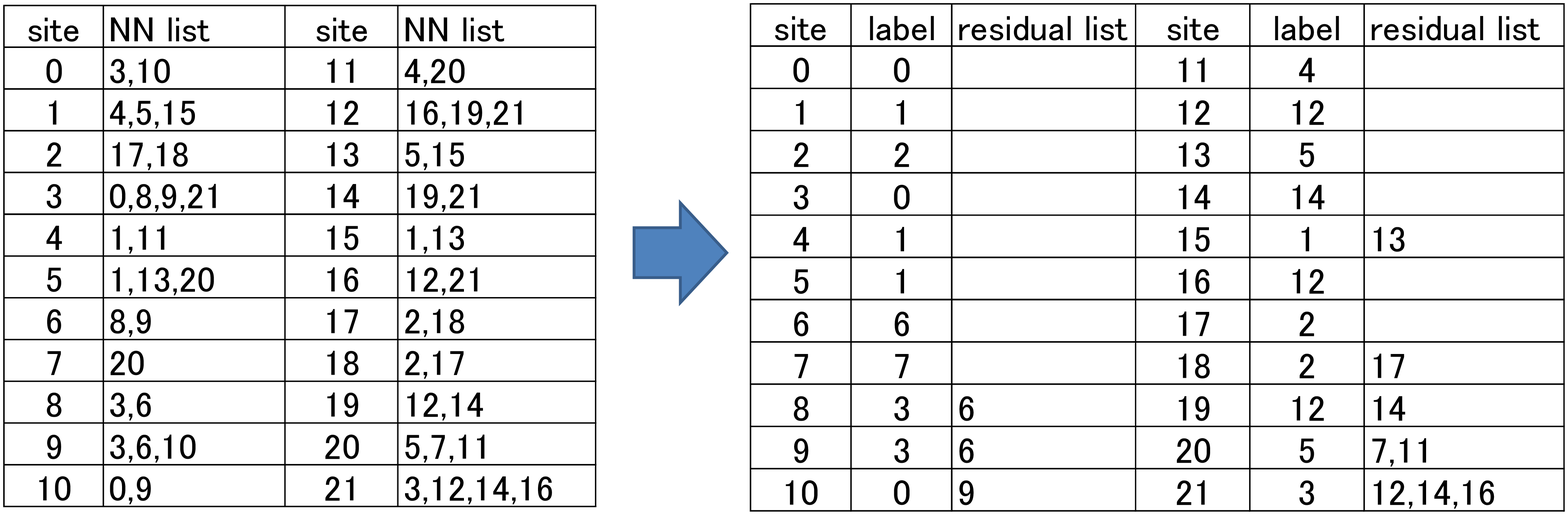}
\caption{\label{fig:method2} 
Initialization step of the generalized algorithm. 
The left panel shows the state 
of the NN list in Fig.~\ref{fig:method1}. 
Each site generates the residual list from the NN list 
by keeping only those neighbors that are numbered lower than the site.
Moreover, each site has had its label set to the lowest value in the residual list, 
 $\textit{label}\mskip 2mu [i]=\textit{min}$, and this number has been removed from the residual list. 
If there is no value in the residual list, each site is labeled with its site number, 
i.e., $\textit{label}\mskip 2mu [i]=i$.
The right panel is the state after the initialization step [step (i)],
}
\end{center}
\end{figure}

\begin{figure}
\begin{center}
\includegraphics[width=12.0cm]{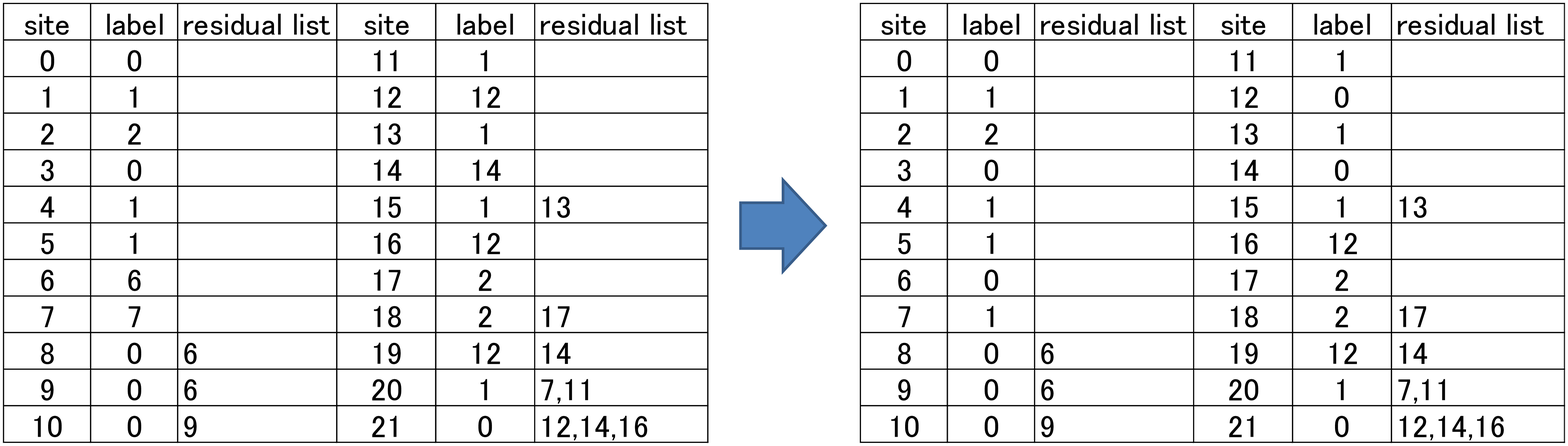}
\caption{\label{fig:method3} 
Label reduction step of the generalized algorithm. 
The left panel 
shows the state after the end of the first analysis step [step (ii)].
The right panel 
shows the state after label reduction [step (iii)],
in which each site is calculated using Algorithm~1.  
}
\end{center}
\end{figure}

\begin{figure}
\begin{center}
\includegraphics[width=12.0cm]{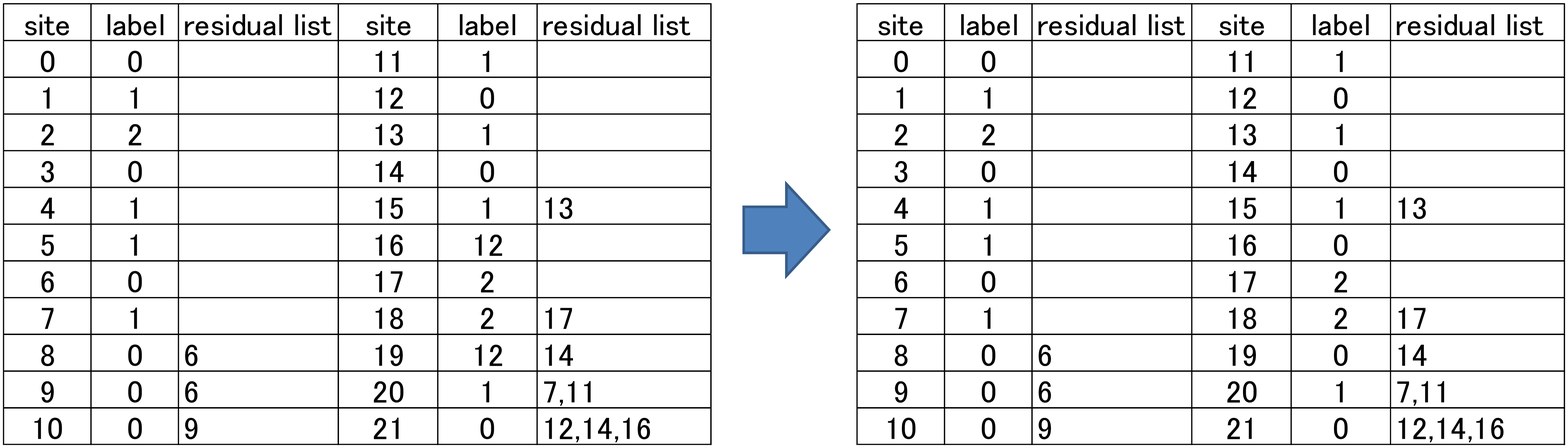}
\caption{\label{fig:method4} 
Final analysis step of the generalized algorithm. 
The left panel 
shows the state after the end of the label reduction step [step (iii)].
The right panel 
shows the state after the second analysis step [step (iv)].  
}
\end{center}
\end{figure}

\section{Generalized GPU-based cluster labeling algorithm}
We now turn to the proposed algorithm. 
The algorithm in \cite{komura15} is specialized to the case of 
square and simple cubic lattices. 
We extend this algorithm of \cite{komura15} to the generalized algorithm. 
The proposed algorithm can be applied to both lattices and to non-lattice 
environments in a uniform fashion as long as a nearest-neighbor (NN) list is supplied. 
Figure~\ref{fig:method1} presents an example of such a list. 
Each site has a list of immediate neighbors.  
In this paper, we prepare the array of NN list in advance. 
The total size of NN list is equal to the product of
the number of sites and the largest size of NN list in all sites, 
and each site gets access to a NN list as $NN\_list[i],NN\_list[i+N],NN\_list[i+2N],...$, where N is the number of sites.

The generalized CCL algorithm also 
consists of the same four steps:
(i) initialization, (ii) analysis, 
(iii) label reduction, and (iv) analysis. 
In the initialization step, 
we generate the residual lists from the NN list by 
keeping only those neighbors that are numbered lower than any given site. 
Moreover, each site has its label set to the lowest site number in the residual list, 
 $\textit{label}\mskip 2mu [i]={min}$, and this is removed from the residual list. 
If there is no number in the residual list, the site is labeled with its site number, 
i.e., $\textit{label}\mskip 2mu [i]=i$.
Figure~\ref{fig:method2} shows an example of this procedure. 
The analysis function tracks the label
from a given site to a new site determined by the value of the label at the given site. 
All sites have $\textit{label}\mskip 2mu [i]=\textit{label}\mskip 2mu [\textit{label}\mskip 2mu [i]]$ calculated until \textit{label} remains unchanged. 
In the label reduction step, 
each site executes Algorithm \ref{alg1}. 
In the label reduction step, 
the calculation at each site uses the pairs 
at each site and the remaining sites in the residual list.  
The label reduction method allows 
each chain of labels, i.e., $\textit{label}\mskip 2mu [\textit{label}\mskip 2mu ]$, to be constructed automatically. 
Figure~\ref{fig:method3} illustrates this procedure. 
In Fig.~\ref{fig:method3}, we show the actual output of our algorithm for the complex network in Fig.~\ref{fig:method1}. 
Finally, each site is assigned the proper cluster label by executing the analysis function again. 
The procedure is illustrated in Fig.~\ref{fig:method4}.

Unlike the CCL algorithms using a single CPU, the label of the cluster is not given serially in this method. 
The label of each cluster depends on the minimum site number in each cluster. 
However, we can renumber the cluster labels to sequence labels 
by using the method shown in Fig. 4 in \cite{komura_multiGPU2}, for example. 

\begin{algorithm}    

\For{$j \in residual\; list$}{
\
// $i$ is a site number\\
{$label\_1 \Leftarrow label[i]$} \;
\While{$label\_1 \neq label[label\_1]$}{
   $label\_1 \Leftarrow label[label\_1]$\;
}\

// $j$ is a number in the residual list of site i\\
{$label\_2 \Leftarrow label[j]$} \;
\While{$label\_2 \neq label[label\_2]$}{
   $label\_2 \Leftarrow label[label\_2]$\;
}\
\\
$flag \Leftarrow true$\;
\If{$label\_1 \neq label\_2 $}{
$flag \Leftarrow false$\; 
}\

\If{label\_1 $<$ label\_2 }{
 $tmp \Leftarrow label\_1$\;
 l$abel\_1 \Leftarrow label\_2$\;
 $label\_2 \Leftarrow tmp$\;
}\

\While{$flag = false$}{
  ${label\_3 \Leftarrow atomicMin(\&label[label\_1], label\_2)}$\;
 \lIf{$label\_3 = label\_2$}{$flag \Leftarrow true$}
 \lElseIf{$label\_3 > label\_2$}{$label\_1 \Leftarrow label\_3$}
 \lElseIf{$label\_3 < label\_2$}{$label\_1 \Leftarrow label\_2; label\_2 \Leftarrow label\_3$}
}\
}\

\caption{ 
Pseudo-code of the label reduction method used in this paper. 
Each site executes this algorithm at the label reduction step [step (iii)]. 
Here, $i$ is a site, $j$ is a nearest-neighbor to site $i$, 
and $label$ is the label at each site. 
\label{alg1} 
}           
\end{algorithm}

\section{Results}
As an application of the generalized CCL algorithm, 
we study bond percolation problems. 
Bond percolation creates bonds between neighboring sites with probability $p$.   
In bond percolation, one considers whether there is a cluster that spans the entire lattice.
If there is such a spanning cluster, the system is regarded as percolating. 
There is a threshold value $p_{c}$ that distinguishes percolating and non-percolating behavior. 
The probability that a spanning cluster is produced is zero for $p<p_c$
and is unity for $p>p_c$ as the system size approaches infinity. 

We test the correctness and performance of the proposed code 
on an NVIDIA TITAN X machine with the CUDA version 7.5 compiler.  
For the random-number generator, we used the XORWOW pseudorandom generator \cite{Xorshift} 
with double precision on the device API in the cuRAND library \cite{curand}. 

\begin{figure}
\begin{center}
\includegraphics[width=8.0cm]{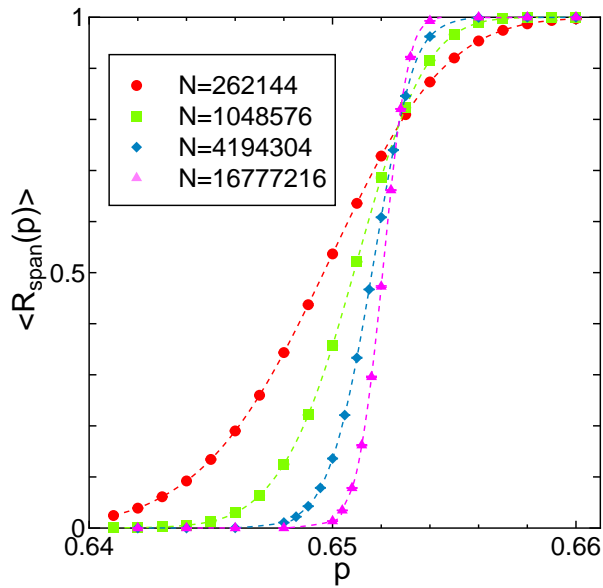}
\caption{\label{fig:probability_honeycomb} 
Spanning probability for bond percolation on the honeycomb lattice.}
\end{center}
\end{figure}

\begin{figure}
\begin{center}
\includegraphics[width=8.0cm]{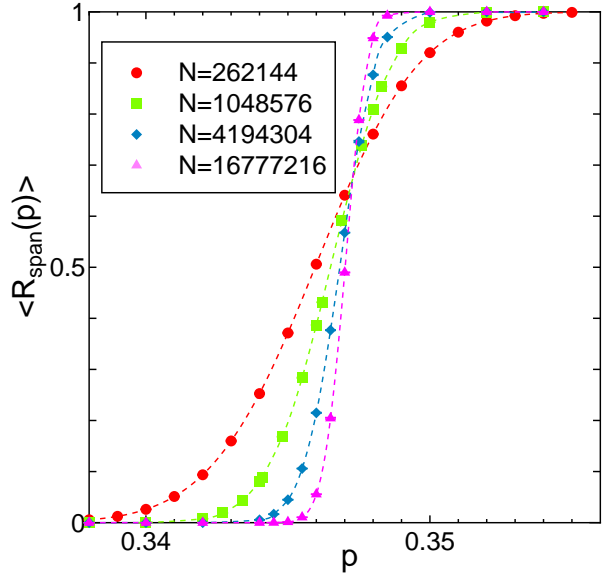}
\caption{\label{fig:probability_triangle} 
Spanning probability for bond percolation on the triangle lattice.}
\end{center}
\end{figure}

\begin{figure}
\begin{center}
\includegraphics[width=8.0cm]{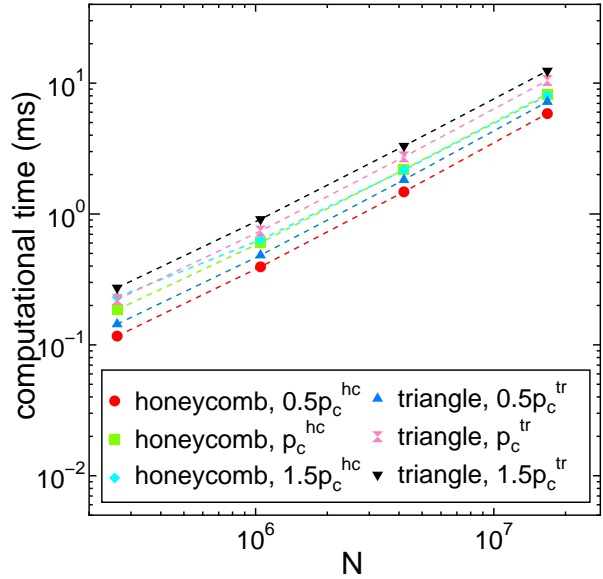}
\caption{\label{fig:time_tri_honey} 
Average computational time for a single realization 
on the honeycomb and triangle lattices 
for probabilities $p= 0.5p_{c}, p_{c}, 1.5p_{c}$. 
We use the threshold value of $p_{c}^{hc} = 1-2\sin(\pi/18)$ for the honeycomb lattice and $p_{c}^{tr} = 2\sin(\pi/18)$ for the triangle lattice. 
}	
\end{center}
\end{figure}

We first deal with bond percolation on the honeycomb and triangle lattices 
to check the correctness of our algorithm. 
For the measured quantity, we use the spanning probability $R_{span}(p)$. 
Figure~\ref{fig:probability_honeycomb} shows the spanning probability $R_{span}(p)$ 
for the honeycomb lattice, and Fig.~\ref{fig:probability_triangle} shows 
the spanning probability for the triangle lattice. 
The total lattice sizes $N$ are $262144$, $1048576$, $4194304$, and $16777216$. 
From these figures, we can see that all curves intersect at one point. 
From the crossing point, we estimate the threshold value $p_{c}^{hc}$ 
on the honeycomb lattice to be $p_{c}^{hc}=0.6527 \pm 0.0002$ and 
the threshold value $p_{c}^{tr}$ 
on the triangle lattice to be $p_{c}^{tr}=0.3472 \pm 0.0002$.  
These estimated values are compatible with the exact values from \cite{Sykes64}, 
namely $p_{c}^{hc} = 1-2\sin(\pi/18)$ 
and $p_{c}^{tr} = 2\sin(\pi/18)$. 
Those results show that our algorithm works well. 

Secondly, we demonstrate the performance for bond percolation on the honeycomb and triangle lattices. 
We give a double logarithmic plot of the average computational times for a single realization 
for the honeycomb and triangle lattices in Fig.~\ref{fig:time_tri_honey}. 
Because the computational time is dependent on the probability $p$,  
we show the average computational times at the probabilities $p= 0.5p_{c}, p_{c}, 1.5p_{c}$ 
on these lattices. We use $p_{c}^{hc} = 1-2\sin(\pi/18)$ as the threshold value for the honeycomb lattice 
 and $p_{c}^{tr} = 2\sin(\pi/18)$ as the threshold value for triangle lattice, 
and the total lattice sizes $N$ are $262144$, $1048576$, $4194304$, and $16777216$.   
From Fig.~\ref{fig:time_tri_honey}, we can see that the average computational time 
is proportional to the total lattice size for all probabilities. 
The computational times on each lattice when the probability is $0.5p_{c}$ is the shortest of those probabilities. 
The computational time for the honeycomb lattice with $N=16777216$ when the probability is $0.5p_{c}^{hc}$ 
is 5.8 ms, and that for the triangle lattice with $N=16777216$ when the probability is $0.5p_{c}^{tr}$ is 7.2 ms. 
Moreover, the computational time for the honeycomb lattice with $N=16777216$ when the probability is $1.5p_{c}^{hc}$ 
is about 1.4 times greater than when the probability is $0.5p_{c}^{hc}$, and  
the computational time for the triangle lattice with $N=16777216$ when the probability is $1.5p_{c}^{tr}$ 
is about 1.7 times greater than when the probability is $0.5p_{c}^{tr}$. 
The differences between the computational times for different probabilities are due to the 
increment in the size of the residual list, and 
the difference between the computational times on the two lattices 
is due to the coordination number.

\begin{figure}
\begin{center}
\includegraphics[width=8.0cm]{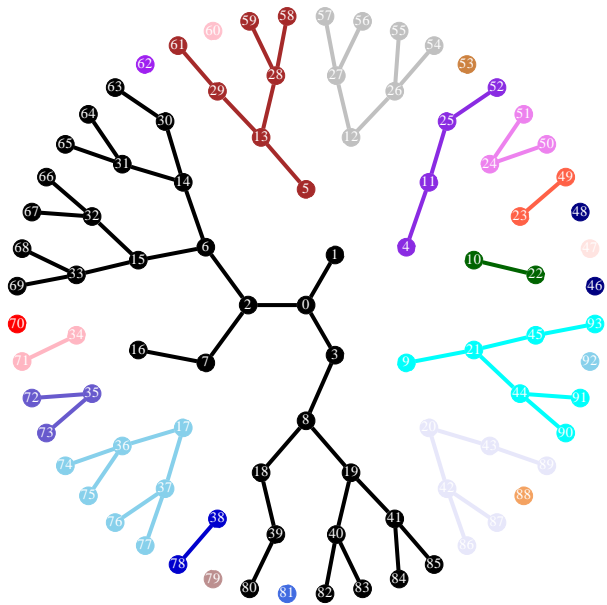}
\caption{\label{fig:bethe} 
An example of one realization of 
the standard position on the Bethe lattice with coordination number $z=3$
when the probability is $p=1.5p_{c}$.
The number at each site is its index, and 
the connection between sites in the same cluster is represented by the same color.
}
\end{center}
\end{figure}

\begin{figure}
\begin{center}
\includegraphics[width=8.0cm]{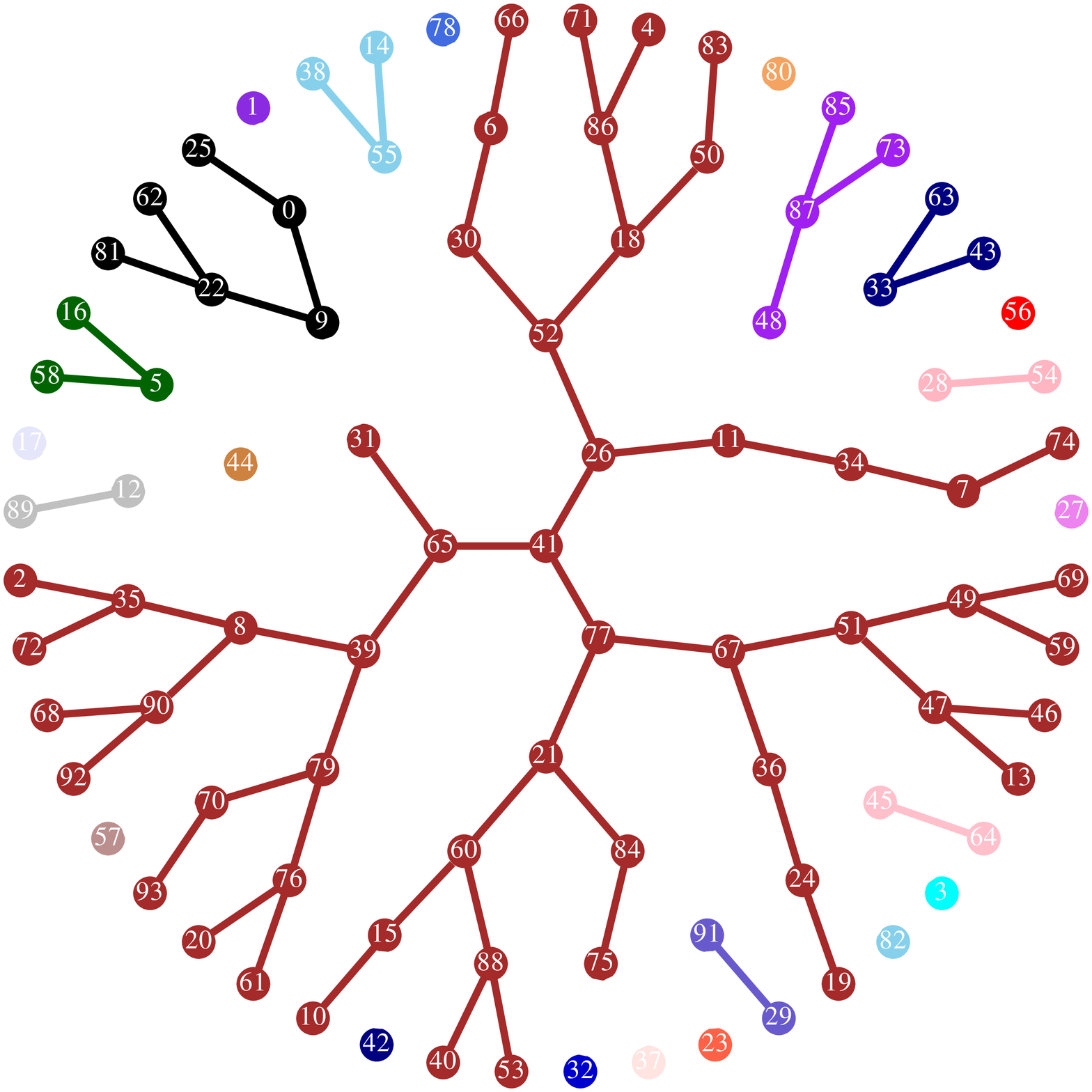}
\caption{\label{fig:bethe_random}
An example of one realization of 
the random position on the Bethe lattice with coordination number $z=3$
when the probability is $p=1.5p_{c}$. 
The number at each site is its index, and 
the connection between sites in the same cluster is represented by the same color.
}
\end{center}
\end{figure}

\begin{figure}
\begin{center}
\includegraphics[width=8.0cm]{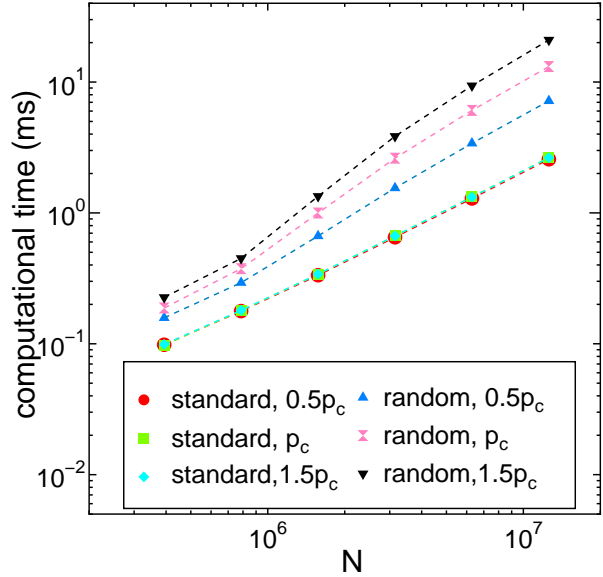}
\caption{\label{fig:time_bethe} 
Average computational times for a single realization 
on the Bethe lattice with coordination number $z=3$
for standard site positions and random site positions 
when the probability is $p= 0.5p_{c}, p_{c}, 1.5p_{c}$.
}
\end{center}
\end{figure}

Next, we deal with bond percolation on the Bethe lattice as a substitute for a network structure. 
The Bethe lattice is a connected cycle-free graph, where each site is connected to $z$ neighbors. 
The threshold value for bond percolation on the Bethe lattice is related to 
the coordination number $z$, and the threshold value of the Bethe lattice with 
coordination number $z$ is $1/(z-1)$. 
We usually give the site number starting from a center site, 
called the standard position in this paper. 
The generalized CCL algorithm is excellent with this numbering method 
because the remaining sites in the residual list for all sites 
become zero at all probabilities.
Thus, we deal with the two Bethe lattices with coordination number $z=3$. 
One is standard position, where the site numbers are given from a center site, 
and the other is random position, where the site numbers are given at random. 
Figure~\ref{fig:bethe} shows an example of one realization of 
standard position on the Bethe lattice with coordination number $z=3$
when the probability is $p=1.5p_{c}$, and 
Fig.~\ref{fig:bethe_random} shows an example of one realization of 
random position on the Bethe lattice with coordination number $z=3$
when the probability is $p=1.5p_{c}$. 
The threshold value of the Bethe lattice with 
coordination number $z=3$ is $p_{c}=1/2$.
Those examples are the results of the actual output.

We examine the computational times of the generalized CCL algorithm 
for bond percolation on the two Bethe lattices.  
Figure~\ref{fig:time_bethe} shows a double logarithmic plot of the average time required for one realization 
on the two Bethe lattices when the probabilities are $p=0.5p_{c}, p_{c}, 1.5p_{c}$. 
We measure the average computational times 
for the total lattice sizes $N =393214, 786430, 1572862, 3145726, 6291454$, and $12582910$. 
In the case of standard position, 
the computational times are almost independent of the probability, 
and the computational time is proportional to the total lattice size. 
In contrast, in the case of random position, 
the computational time is dependent on the probability. 
The computational times of the random position for the total lattice sizes $N\leq786430$  
are proportional to the total lattice sizes for all probabilities, and 
those for the total lattice sizes $N\geq786430$ deviate from being proportional to the total lattice size at all probabilities. 
This change is related to 
the amount of L2 cache, which is 3MB in the TITAN X.
The computational time is shortest when the probability is $0.5p_{c}$. 
The computational time for random position with $N=12582910$ when the probability is $0.5p_{c}$ is 7.1 ms, 
and that when the probability is $p= 1.5p_{c}$ is 21 ms. 
Moreover, the computational time for random position with $N=12582910$ when the probability is $1.5p_{c}$  
is about 8.0 times greater than that for standard position with $N=12582910$. 
Because we give the site number at random, 
the remaining sites in the residual list for each site are also random. 
This randomness causes a load imbalance.  
Thus, the computational times for random position are greater than those for standard position. 
However, those results show that 
the generalized CCL algorithm can be applied directly to any structure without reordering.

\section{Summary and discussion}
We have proposed a generalized GPU-based CCL algorithm 
that can be applied to both various lattices and to non-lattice environments. 
Because the algorithm in \cite{komura15} is specialized to the case of 
square and simple cubic lattices,  
we have extended our previous labeling algorithm \cite{komura15}, 
which does not use conventional iteration, to the generalized method. 
The proposed algorithm can be applied directly to any structure without reordering. 

We chose the bond percolation problem as a test application.
We first dealt with bond percolation on the 
honeycomb and triangle lattices to confirm the correctness of our algorithm. 
We used the spanning probability as a measured quantity, and    
reproduced some well-known results. 
Secondly, we showed the performance for bond percolation on the honeycomb and triangle lattices. 
Because the computational time is dependent on the probability $p$,  
we reported the average computational times for the probabilities $p= 0.5p_{c}, p_{c}, 1.5p_{c}$ 
The computational time for the honeycomb lattice with $N=16777216$ when the probability is $0.5p_{c}$ 
is 5.8 ms, and that for the triangle lattice with $N=16777216$ when the probability is $0.5p_{c}$ is 7.2 ms. 
Finally, we dealt with bond percolation on the Bethe lattice as a substitute for a network structure.
Because the remaining sites in the residual list for all sites become zero at all probabilities
when we give the site numbers starting from a center site, we dealt with the 
two Bethe lattices with coordination number $z=3$.  
The computational times for random position, where the site numbers are given at random, 
are greater than those for standard position, where the site numbers are given from a center site, at all probabilities. 
However, the generalized CCL algorithm can be applied directly to any structure without reordering.

We finally emphasize the efficiency of the proposed algorithm. 
This algorithm is very powerful; it can be adapted both to 
a variety of lattices and to non-lattice environments in a uniform fashion 
as long as an NN list is supplied. 
That is, this algorithm can be adapted to any structure 
by simply replacing the NN list. 
Moreover, this algorithm is suitable for the GPU architecture which 
the memory access latency can be hidden with calculations instead of big data caches.
This algorithm is realized by indirect memory access and random memory access, 
and frequently cause the simultaneous update with atomic function.
Thus, this algorithm is unsuitable for the architecture of 
traditional multiprocessors because this algorithm can not be vectorized
and frequently cause the cache coherence.
We hope that many researchers show more interest in developing parallel algorithms 
from a standpoint of architecture.

\section*{Acknowledgments}
This work was supported by KAKENHI grant 15K21623 from the Japan Society for the Promotion of Science.

\end{document}